# Field-induced transition from even to odd parity superconductivity in CeRh$_2$As$_2$

S. Khim,[1†*] J. F. Landaeta,[1†] J. Banda,[1] N. Bannor,[1] M. Brando,[1] P. M. R. Brydon,[2] D. Hafner,[1] R. Küchler,[1] R. Cardoso–Gil,[1] U. Stockert,[1] A. P. Mackenzie,[1,3] D. F. Agterberg,[4] C. Geibel,[1] and E. Hassinger[1,5*]

[1]Max Planck Institute for Chemical Physics of Solids, 01187 Dresden, Germany
[2]Department of Physics and MacDiarmid Institute for Advanced Materials and Nanotechnology, University of Otago, P.O. Box 56, Dunedin 9054, New Zealand
[3]Scottish Universities Physics Alliance, School of Physics and Astronomy, University of St Andrews, St Andrews KY16 9SS, United Kingdom
[4]Department of Physics, University of Wisconsin–Milwaukee, Milwaukee, Wisconsin 53201, USA
[5]Technical University Munich, Physics department, 85748 Garching, Germany

[†]These authors contributed equally to the research.
[*]Correspondence should be addressed to elena.hassinger@cpfs.mpg.de and seunghyun.khim@cpfs.mpg.de

**We report the discovery of two-phase unconventional superconductivity in CeRh$_2$As$_2$. Using thermodynamic probes, we establish that the superconducting critical field of its high-field phase is as high as 14 T, remarkable in a material whose transition temperature is 0.26 K. Furthermore, a *c*-axis field drives a transition between two different superconducting phases. In spite of the fact that CeRh$_2$As$_2$ is globally centrosymmetric, we show that local inversion symmetry breaking at the Ce sites enables Rashba spin-orbit coupling to play a key role in the underlying physics. More detailed analysis identifies the transition from the low- to high-field states to be associated with one between states of even and odd parity.**

# Introduction

A major advance in condensed matter physics over the past four decades has been the discovery of new forms of so-called unconventional superconductivity involving exotic, purely electronic pairing. Throughout this process, a surprise has been how many of the new materials have simple, single-component phase diagrams. Based on knowledge of superfluidity in $^3$He (*1*) and on the fact that degeneracies or near-degeneracies can be expected to result from many of the electronic mechanisms for unconventional superconductivity (*2*), one might have expected widespread observation of rich temperature - magnetic field phase diagrams with transitions within the superconducting state between different superconducting order parameters. Until now, however, the only stoichiometric superconductor that has been well established to have such rich temperature - magnetic field superconducting phase diagram at ambient pressure is UPt$_3$ (*3–5*).

In this paper we report the discovery of a second such example, in the heavy fermion material CeRh$_2$As$_2$. Experimentally, we show that it has extremely high superconducting critical fields of up to 14 T in spite of a superconducting transition temperature $T_c$ of only 0.26 K. Further, when the magnetic field is applied along the crystallographic *c*-axis, the superconducting state contains a well-defined internal phase transition at approximately 4 T that we identify using several thermodynamic probes. We also show that these remarkable observations do not result from similar underlying physics to that at play in UPt$_3$. Rather, the key superconducting properties of CeRh$_2$As$_2$ are a manifestation of local inversion symmetry breaking and consequent Rashba spin-orbit coupling in an overall inversion-symmetric crystal structure (*6–10*). The possibility that this kind of physics might drive multi-phase superconductivity has been considered in the theoretical literature (*11–14*), but its clear observation is, to our knowledge, completely new. Combined with intriguing normal state physics that likely also results from the unusual crystalline environment of Ce, our observations ensure that CeRh$_2$As$_2$ will be a benchmark material in which to study the influence of spin-orbit coupling on electronic mechanisms for unconventional superconductivity.

# Results

CeRh$_2$As$_2$ crystallizes in the centrosymmetric tetragonal CaBe$_2$Ge$_2$-type structure (*16*) (Fig. 1A) in which Ce is alternatively stacked with two different Rh-As blocks along the *c*-axis; Rh(1) (As(1)) is tetrahedrally coordinated by As(2) (Rh(2)), respectively. Hence, there are two Ce sites per unit cell and each Ce site lacks inversion symmetry locally with the local polar $C_{4v}$ point group. The lattice inversion point lies in the middle of the line connecting those two Ce sites. We believe this intriguing structure plays a central role in the physics of the superconducting state.

The high-temperature magnetization of single-crystalline CeRh$_2$As$_2$ shows paramagnetic Curie-Weiss behavior with an effective moment of 2.56 $\mu_B$ per Ce, corresponding to a Ce$^{3+}$ valence state (Fig. 1B). In the whole temperature range, the $ab$-plane magnetization is larger than the $c$-axis one, by up to a factor of two at low temperature. The resistivity $\rho(T)$ depicted in Fig. 1C displays typical heavy fermion behavior with increasing resistivity upon decreasing temperature due to the Kondo effect. At temperatures below a characteristic local maximum at approximately 40 K, $\rho(T)$ decreases when the heavy quasiparticle bands are formed by hybridization of local 4$f$ electrons with the conduction electrons. The large thermopower $S(T)$ is typical for a Kondo lattice system (Fig. 1D). Below 4K the specific heat $C(T)/T$ in Fig. 1E (where the nuclear contribution has been removed (*15*)) increases towards low temperature following a power law with $C/T \propto T^{-0.6}$, suggesting non-Fermi liquid behavior and proximity to a quantum critical point (*17*). It reaches a large value of 1 J/molK$^2$ at $T = 0.5$ K. The Kondo temperature in CeRh$_2$As$_2$ is between 20 K and 40 K, as estimated from the magnetic entropy $S_{mag}(T)$ shown in Fig. 1F (see the supplemental material (SM) (*15*)). Interestingly, $S_{mag}$ monotonically increases to reach the value $R \ln 4$ without a signature of a plateau at $R \ln 2$, where $R$ is the ideal gas constant, suggesting that the two low-lying doublets of the crystal electric field (CEF) are very close in energy. The estimated separation of ~ 30 K that is comparable with the Kondo energy could lead to a possible quasi-quartet ground state (*15*, *18*). This is a rare example among the tetragonal Ce systems which usually exhibit a separation of ≥ 100 K and again highlights the unusual local Ce environment in CeRh$_2$As$_2$.

Below 1 K, two anomalies appear in the specific heat as shown in Fig. 1E and G. A small hump is visible at $T_0 \approx 0.4$ K where the data depart from the power-law behavior extrapolated from high temperatures which is depicted by the dashed line. It hints at a phase transition to an ordered state. The large jump below 0.3 K results from the transition to a superconducting state involving the $f$ electrons. An equal entropy analysis reveals $T_c = 0.26$ K and a height of the jump at $T_c$ of $\Delta C/C|_{Tc}$ ≈ 1, of the same order of magnitude as the Bardeen-Cooper-Schrieffer (BCS) value of ~ 1.4. The residual Sommerfeld coefficient $\gamma = C/T$ for $T \rightarrow 0$ is probably a sign of impurities. The diamagnetic drop of the ac-susceptibility confirms entry to the superconducting state (Fig. 1H) at a similar $T_c$ for the transition midpoint but a slightly higher onset temperature while the first drop in resistivity takes place at 0.39 K (Fig. 1I). Although this is close to $T_0$ in zero field, the increase of $T_0$ with in-plane fields (see the specific heat data at 8 T and 12 T in Fig. 2A) shows that $T_0$ is not associated with superconductivity but likely signals some other kind of order. Its origin is yet to be determined, but the absence of an anomaly in the magnetic susceptibility at $T_0$ suggests that it might have Ce-4$f$ multipolar or nematic character. We ascribe the higher $T_c$ in the resistivity and susceptibility to inhomogeneity in the material as known from other heavy fermion systems (*19, 20*). As a summary of these first results, CeRh$_2$As$_2$ is a heavy fermion superconductor where the lowest CEF levels are separated by an energy of similar size as the Kondo temperature, both of the order of 30 K. Just before becoming superconducting at low temperature the system enters a state of unknown origin. For the remainder of this paper, we focus on the extraordinary superconducting properties as established experimentally using magnetic susceptibility and thermodynamic probes.

In Fig. 2 we show the temperature dependence of the specific heat $C/T$ (panels A and B) and the magnetic ac-susceptibility $\chi_{ac}$ (panels C and D) for different magnetic fields between 0 T and 14 T for $H \parallel ab$ (A and C) and $H \parallel c$ (B and D). $T_c$ is defined via the equal entropy method in $C/T$ and at the onset of the susceptibility transition (chosen arbitrarily by the temperature where $\chi_{ac}$ has dropped to the value indicated by the dashed line). It shifts down with increasing field. In $\chi_{ac}$ we observe a relatively strong shift of $T_c$ in a field of 100 mT that is absent in the specific heat, again a sign of non-bulk superconductivity (*15*). Increasing the field further reduces $T_c$ more slowly. The superconducting transition is completely suppressed down to 0.05 K at magnetic fields of 14 T for $H \parallel c$ and 2 T for $H \parallel ab$. We note that, especially for $H \parallel c$, these are remarkably large critical fields for a superconductor with at $T_c$ of only 0.26 K. For $H \parallel c$ the temperature sweeps (Fig. 2B and D) imply a kink in the $T_c(H)$ curve where above 4 T, the decrease of $T_c$ is slower than below this field.

A pronounced kink in $T_c(H)$ is suggestive of the existence of two superconducting phases. Indeed, this is confirmed by field sweeps of the ac-susceptibility and two separate thermodynamic probes, magnetization and magnetostriction (Fig. 3A-C). Remarkably, all three provide striking evidence of a phase transition. Below $T = 0.2$ K, pronounced kinks in all three observables are seen at a characteristic field, $H^* = 3.8$ T, that is almost temperature independent. As shown in Fig. 3A and the inset of Fig. 3C, diamagnetic shielding and zero resistivity persist throughout the entire region of interest, further proving that this is a phase transition within the superconducting state.

Using the values of $T_c$, $H_{c2}$ (defined in $\chi_{ac}(T)$ at the onset in the same way as in the temperature sweeps) and $H^*$ from our measurements, we show the superconducting phase diagrams of CeRh$_2$As$_2$ for out-of-plane and in-plane fields in Fig. 4A and B, respectively. From these phase diagrams the superconducting critical field can be extrapolated to $H_{c2}(0) \approx 14$ T for $H \parallel c$ and 1.9 T for $H \parallel ab$. For $H \parallel c$ two superconducting states appear, labelled as SC1 and SC2, separated by a line that intersects the strong kink in the $H_{c2}(T)$ curve in a multicritical point.

It is useful to estimate the upper critical fields with the Werthamer-Helfand-Hohenberg (WHH) formula $H_{orb} \approx 0.693 T_c (-dH_{c2}/dT)_{Tc}$ which only uses parameters near $T_c$ where Pauli paramagnetic pair-breaking effects are parametrically suppressed (*21*). Using the large experimental slopes ($-dH_{c2}/dT)_{Tc} = 97$ T/K for $H \parallel c$ and ($-dH_{c2}/dT)_{Tc} = 45$ T/K for $H \parallel ab$, this yields $H_{orb} \approx 17$ T and $H_{orb} \approx 8$ T, respectively. Their anisotropy of a factor of $\approx 2$ reflects the anisotropy of the effective mass since $H_{orb} \propto m^{*2}$ (*22*). The corresponding BCS coherence lengths $\xi = (\Phi_0 / 2\pi H_{c2}(T=0))^{0.5}$ are accordingly small, lying below 100 Å. These estimates suggest that the upper critical field of SC2 along the *c*-axis is not Pauli-paramagnetically suppressed. In contrast, we find that the superconducting state SC1 is strongly Pauli limited with Pauli critical fields that are enhanced compared to the Clogston-Chandrasekhar limit ($H_P \approx 0.5$ T) and 3 times larger for $H \parallel c$ than for $H \parallel ab$. This factor represents the scaling factor of the experimental critical field of SC1 for the two magnetic field directions. The enhancement and the anisotropy result from a combination of correlations, anisotropic renormalized *g*-factors and enhanced spin susceptibilities in the superconducting state (*13*, *22*, *23*). All of these are reminiscent of the role of strong Rashba spin-orbit coupling on superconductivity (*24*, *25*). Such a coupling is often assumed to require inversion symmetry to be broken, which is not the case in CeRh$_2$As$_2$. However, as has been pointed out a number of times (*6–13*, *26*, *27*), a Rashba term can exist due to local inversion-symmetry breaking even in an overall inversion-symmetric structure, and influence the superconductivity. As we now show, this Rashba term is large in CeRh$_2$As$_2$, and holds the key to understanding the phase diagram shown in Figs. 4A and B.

## Discussion

To illustrate how spin-orbit coupling is determining the key physics of CeRh$_2$As$_2$ in spite of the presence of global inversion symmetry, we develop a model taking into account the unusual features of its structure. In particular, noting that the Ce 4$f$ electrons are key to the heavy quasiparticle bands that give rise to superconductivity, we consider Wannier functions for these bands that are centered on the Ce sites. The Ce atoms sit at sites with a local $C_{4v}$ symmetry, for which electronic states belong to CEF doublets of either $\Gamma_6$ or $\Gamma_7$ symmetry. It is possible to write down a symmetry-based tight binding Hamiltonian, which takes the same form for Wannier functions belonging to either of these two doublets

$$\begin{aligned} H_N = {} & t_1 [\cos(k_x) + \cos(k_y)] - \mu + \alpha_R \tau_z [\sin(k_x)\sigma_y - \sin(k_y)\sigma_x] \\ & + t_{c,1} \tau_x \cos\left(\tfrac{k_z}{2}\right)\cos\left(\tfrac{k_x}{2}\right)\cos\left(\tfrac{k_y}{2}\right) + t_{c,2} \tau_y \sin\left(\tfrac{k_z}{2}\right)\cos\left(\tfrac{k_x}{2}\right)\cos\left(\tfrac{k_y}{2}\right) \\ & + \lambda \tau_z \sigma_z \sin k_z (\cos k_x - \cos k_y) \sin k_x \sin k_y. \end{aligned} \quad (1)$$

In this normal-state Hamiltonian $H_N$, the $\sigma_i$ Pauli matrices represent the two Kramer's spin-like degenerate states of the $\Gamma_6$ or $\Gamma_7$ doublets and the $\tau_i$ Pauli matrices represent the two Ce site degrees of freedom in each unit cell. Importantly, since these two Ce site degrees of freedom are related by inversion symmetry, the $\tau_z$ matrix is odd under inversion symmetry (this follows because this matrix changes sign when the two Ce sites are interchanged). This Hamiltonian assumes that only one CEF doublet needs to be included per Ce site, which is reasonable given that the energy difference between two local CEF doublets has been observed to be about 30 K and superconductivity sets in at 0.26 K. Eq. 1 reveals that a Rashba-like spin-orbit interaction, denoted by the constant $\alpha_R$, is allowed by symmetry, in spite of the overall inversion symmetric lattice. This term appears because of the presence of the odd inversion $\tau_z$ operator which encodes the fact that the Ce sites are not sites of inversion symmetry. Eq. 1 also reveals an additional Ising-like spin-orbit coupling term denoted by $\lambda$, with a $\tau_z \sigma_z$ dependence. This term will be much smaller than $\alpha_R$ since $\alpha_R$ originates from nearest neighbor ($a$, 0, 0)-type hoppings while $\lambda$ requires much longer range ($a$, 2$a$, $c$)-type hoppings, for this reason we will set $\lambda = 0$ in the following. The two parameters $\tau_{c,i}$ correspond to $c$-axis ($a$/2, $a$/2, $c$/2) hoppings between the two Ce sublattice sites.

Now we turn to the superconducting state. Density functional theory (DFT) reveals that the band structure for the conduction electrons is quasi two-dimensional (*28*), so it is natural to assume that the quasi-particle interactions that give rise to superconductivity originate in the two-dimensional square Ce layers, which consist of only Ce sites from the same sublattice. So, in terms of $\tau_i$ operators, the Cooper pairs can then have only a $\tau_0$ or a $\tau_z$ dependence. Formally, $\tau_0$ ($\tau_z$) describes Cooper pair wavefunctions that have the same (opposite) sign on the two Ce sublattice sites. For simplicity, we will assume that each sublattice prefers a spin-singlet *s*-wave Cooper pair. This is not essential for the arguments presented below which rely on $\tau_i$ structure of the Cooper pairs. Below we explicitly consider the even parity gap function $\Delta_e = \Delta_{\tau 0}$ and the odd parity gap function $\Delta_o = \Delta_{\tau z}$. We note that similar gap functions have been discussed in three-dimensional $Cu_xBi_2Se_3$ (*29*) and, remarkably, $\Delta_o$-type gap functions were originally proposed by P. W. Anderson as the generic form of odd-parity superconductivity in heavy fermion materials (*11*). These results indicate that $\Delta_e$ and $\Delta_o$-type gap functions are more generally stable than the quasi two-dimensional limit we consider here.

Eq. 1 gives rise to two bands each with eigenstates that can be characterized by pseudospin (*30*, *31*). It is instructive to express the gap functions in terms of this pseudospin basis. The gap function $\Delta_e$ becomes a usual pseudospin-singlet state, while $\Delta_o$ becomes a pseudospin-triplet gap function (*32*),

$$\Delta \tau_z \to \sin k_y\, s_x - \sin k_x\, s_y \qquad (2)$$

where the $s_i$ Pauli matrices denote the pseudospin-triplet gap degrees of freedom. The reason for the existence of this pseudospin-triplet state stems from the odd-inversion symmetry of the $\tau_z$ operator, which ensures that $\Delta_o$ has odd parity, and must therefore be a pseudospin triplet. Importantly, this pseudospin-triplet gap function describes Cooper pairs with a pseudospin projection along the *c*-axis; crucially this is parallel to the pseudospin polarization induced by a *c*-axis magnetic field, and so this state is not Pauli limited. However, $\Delta_e$ will be Pauli suppressed, and this difference underlies the origin of the field-induced transition. For in-plane fields, both $\Delta_o$ and $\Delta_e$ are Pauli suppressed, so the even parity solution wins out at all temperatures and fields. Therefore, we identify SC1 with the even parity state $\Delta_e$ and SC2 with the odd parity state $\Delta_o$. As described in more detail in Ref. (*32*) this result is generic, providing a robust qualitative explanation for the observed phase transition in *c*-axis fields, independent of microscopic details of the eventual pairing mechanism.

To carry out a semi-quantitative analysis, we expand $H_N$ near the $\Gamma$ point and take $t_c = t_{c,1} = t_{c,2}$. This then yields a simplified model with a cylindrical Fermi surface (with momentum $k_F$) that is defined by the parameter $\tilde{\alpha} = \alpha_R k_F / t_c$. This simplified model closely resembles that of Ref. (*12, 13*), which finds two transitions under *c*-axis fields. In the limit that the Zeeman field does not appreciably change the band structure, some useful results can be found. The first gives the ratio of the pairing interactions for the odd ($V_o$) and even ($V_e$) gap functions

$$\frac{V_o}{V_e} = \frac{\tilde{\alpha}^2}{1 + \tilde{\alpha}^2} \tag{3}$$

revealing that the *c*-axis hopping terms suppress $\Delta_o$, so that $\Delta_e$ is the stable state in zero field. The second useful result gives the Pauli limiting field for the $\Delta_e$ state when the field is applied along the *c*-axis

$$\frac{H_P^c}{H_P} = \sqrt{1 + \tilde{\alpha}^2} \tag{4}$$

and reveals that $\alpha_R$ enhances the *c*-axis Pauli critical field for $\Delta_e$, explaining why the low-field phase survives to *c*-axis fields that are notably larger than the Clogston-Chandrasekhar field. We find that there is also a smaller enhancement for fields in-plane, but this enhancement cannot be expressed analytically. Using this simplified model and including an orbital critical field, we fit the experimental results as shown in Fig. 4C and D. In particular, using the estimates for upper critical fields $H_{\text{orb}}$ and the *g*-factors (independently obtained from the CEF fits and from the Wilson ratio (*15*)), we fit the $\Delta_e$ state for in-plane fields and obtain $\tilde{\alpha} = 2.5$ and a Maki-parameter $\alpha_M = 6.8$ (Fig. 4D). The value of $\alpha_M$ agrees with an estimation from normal state properties (*15*). These two parameters are all that are needed to determine the *c*-axis upper critical field for $\Delta_e$ in Fig. 4D. The upper critical field shown for $\Delta_o$ is purely orbital limited and is found here by taking $\Delta_o/\Delta_e = 0.93$. These fits suggest that the situation in CeRh$_2$As$_2$ is in remarkably close correspondence with that described in earlier theoretical work on a model for which, until now, there were no candidate materials (*12, 13*). It is remarkable that such a simple theory captures the key features of the magnetic field - temperature phase diagram for both in-plane and out-of-plane magnetic fields, especially in a heavy fermion system.

As a final point, we look at the multicritical point in the phase diagram for $H \parallel c$. Thermodynamic considerations forbid that three second-order transition lines meet at a multicritical point (*33*, *34*). However, the phase diagram as experimentally determined here for $H \parallel c$ is thermodynamically possible if the lines separating the SC1 and SC2 from the normal state are second order and the line separating the two superconducting states is first order (*34*). The sharpness of the transition at $H^*$ in all measurements and the fact that the transition line becomes horizontal for $T \rightarrow 0$ support the idea that this transition is of first order. Although we have not observed hysteretic behavior in our experiments, weak hysteresis frequently falls below experimental resolution, and it would be dangerous to read too much physics into a failure to resolve it. From the slopes of the transition lines near the multicritical point, we can estimate the change of the specific heat jump below and above the kink field (see Suppl. (*15*)). We find $\Delta C(H < H_{\text{kink}}) \approx (0.85 \pm 0.2) \, \Delta C(H > H_{\text{kink}})$, in agreement with the increase of the jump observed in experiment.

We cannot exclude the fact that the phase diagram including the normal state is more complicated. The putative ordered state below $T_0 \approx 0.4$ K also seems to be suppressed at $H \approx 4$ T (see Fig. 2B) and the transition line might join the multicritical point as a fourth transition line, allowing the transition within the superconducting state to be second order. This scenario is also thermodynamically possible but would place further constraints on the slopes of the lines and the ratios of the specific heat jumps. A more detailed study of the specific heat would, therefore, likely be able to resolve the issue, but that is beyond the scope of the present paper. It is intriguing to compare our findings with exciting recent developments in UTe$_2$ (*35*, *36*). There, multi-phase superconductivity has been established in the magnetic field - temperature phase diagram under the application of hydrostatic pressure (*37*). A substantial body of theoretical work has been done on UTe$_2$ (*38–42*). In contrast to our ambient pressure findings on CeRh$_2$As$_2$, the majority opinion is that the relevant phases are all triplet and spin fluctuations, not Rashba spin-orbit coupling, are thought to be the main driver of the relevant physics. While we do not claim that the simple theory outlined above is a unique explanation for our observations, we note that it accounts for key features of the magnetic field - temperature phase diagram and naturally predicts a first-order transition between the two superconducting phases. We believe that it is at the very least a useful starting framework within which to identify the key physics.

## Summary and conclusion

In conclusion, we have identified $CeRh_2As_2$ as one of very few unconventional superconductors for which a multi-phase superconductivity has been established using thermodynamic probes. The root cause of this intriguing behaviour seems to be the unusual crystalline environment of the Ce atoms whose *f* electrons participate in the Fermi sea at low temperatures. A Rashba spin-orbit coupling that is allowed by the lack of local inversion symmetry (even though the global inversion is respected) has a profound effect on the rich superconducting physics. Many open questions remain about the precise nature of the superconducting mechanism, including the possibility that the superconductivity in zero applied magnetic field condenses from a non-superconducting state that already includes unidentified order. Like the superconductivity, that 'hidden' order is probably rooted in the unusual Ce environment. $CeRh_2As_2$ therefore opens new avenues of research into the importance of local symmetry breaking not just on superconductivity, but on metallic correlated electron order as well.

## Acknowledgments

We strongly appreciated discussions with David Cavanagh, Onur Erten, Mark Fischer, Jun Sung Kim, David Möckli, Aline Ramirez, Burkhard Schmidt, Manfred Sigrist, Peter Thalmeier, Yoichi Yanase, Gertrud Zwicknagl, and SteffenWirth. We thank Markus König, Ulrich Burkhardt, Monika Eckert, and Sylvia Kostmann for EDX measurements. We acknowledge funding of the Physics of Quantum Materials department and the research group "Physics of unconventional metals and superconductors (PUMAS)" by the Max Planck Society. C.G. and E.H. acknowledge support from the DFG through grant GE 602/4-1 Fermi-NESt. P.M.R.B. was supported by the Marsden Fund Council from Government funding, managed by Royal Society Te Apārangi.

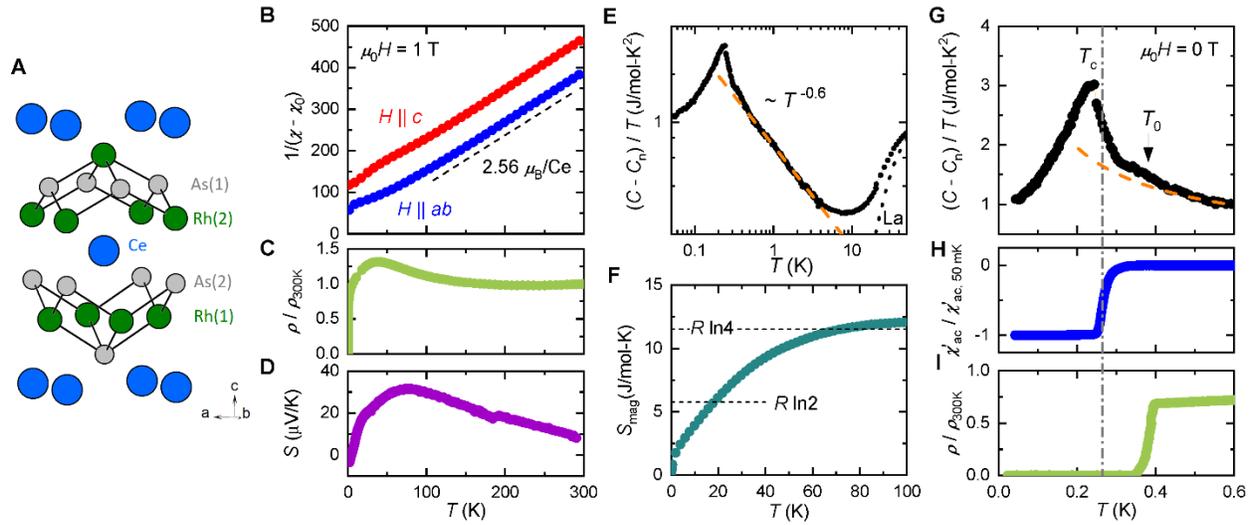

Figure 1: **Crystal structure and heavy fermion superconductivity in CeRh$_2$As$_2$**. The magnetic field $\mu_0 H = 0$ unless indicated otherwise. (**A**) Crystal structure of CeRh$_2$As$_2$. (**B**) Inverse magnetic susceptibility $\chi(T)$ (after subtracting a temperature ($T$)-independent contribution $\chi_0$) in $\mu_0 H = 1$ T applied in the *ab*-plane and along the *c*-axis. (**C**) The resistivity $\rho(T)$ with the current in the *ab*-plane, normalised at 300 K. (**D**) The thermopower $S(T)$ with a temperature gradient in the *ab*-plane. (**E**) The specific heat $C/T(T)$. A nuclear contribution $C_n$ was subtracted at low $T$ (*15*). The dotted line presents the LaRh$_2$As$_2$ data used to remove the phonon contribution. The dashed line represents the power-law $T$-dependence. (**F**) The Ce magnetic entropy $S_{mag}(T)$. (**G-I**) Experimental signatures at the superconducting transition $T_c$ and at the transition $T_0$, see the text for details. (**G**) Specific heat $(C - C_n)/T(T)$ including the same dashed line as in (**E**) and transition temperatures as indicated. (**H**) Normalised ac-susceptibility $\chi'_{ac}$. (**I**) The normalised electric resistivity $\rho(T)$.

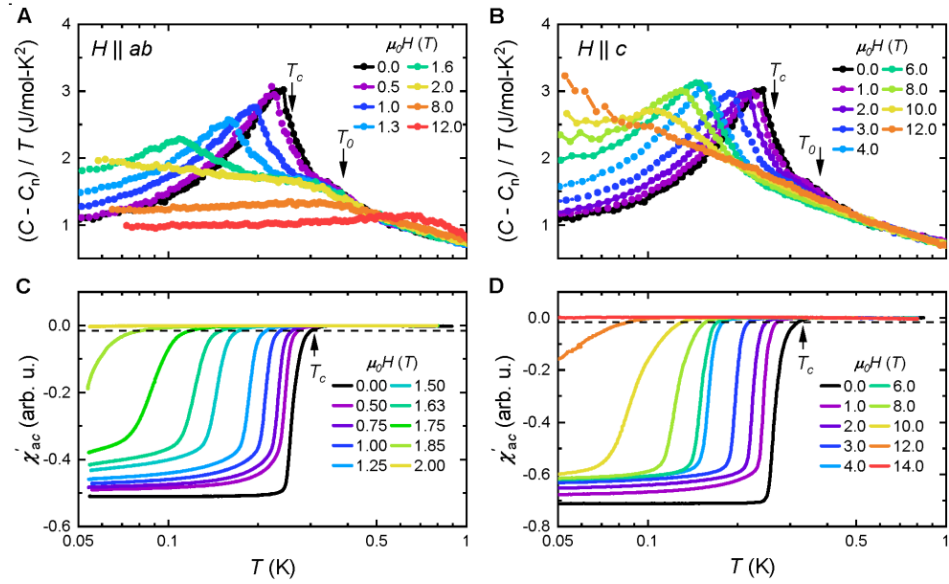

Figure 2: **Evolution of the superconducting transition with magnetic fields**. Temperature dependence of the specific heat $C/T$ (**A**) and the real part of the ac-susceptibility $\chi'_{ac}$ (**C**), respectively, for $H \parallel ab$. (**B**) and (**D**) same for $H \parallel c$. The dashed line in (**C**) and (**D**) indicates the value of $\chi'_{ac}$ where the onset temperature $T_c$ is defined.

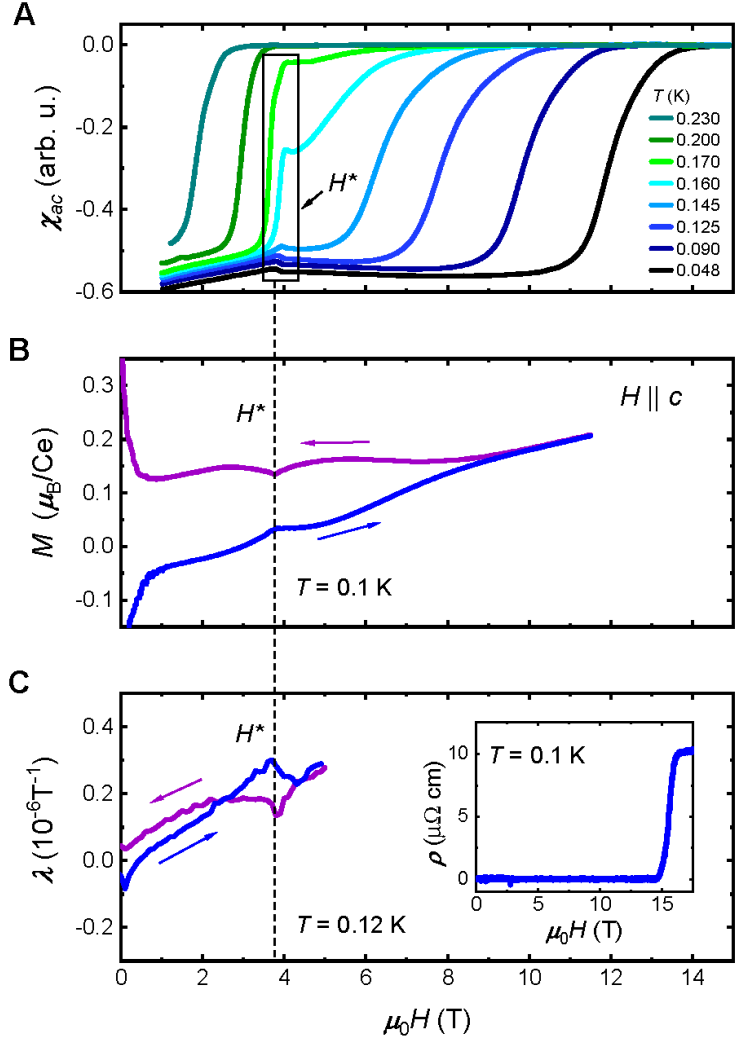

Figure 3: **Phase transition inside the superconducting state for $H \parallel c$.** (**A**) shows the absolute value of the magnetic susceptibility $\chi_{ac}$ for different temperatures as indicated. In the rectangle we highlight the transition inside the superconducting state near $H^* = 3.8$ T present for all temperatures below 0.2 K. (**B**) shows the magnetization $M$ at 100 mK. (**C**) The magnetostriction at 0.12 K. The inset depicts the resistivity measurements at 0.1 K. In (**B**) and (**C**) the $H^*$ transition also appears at 3.8 T.

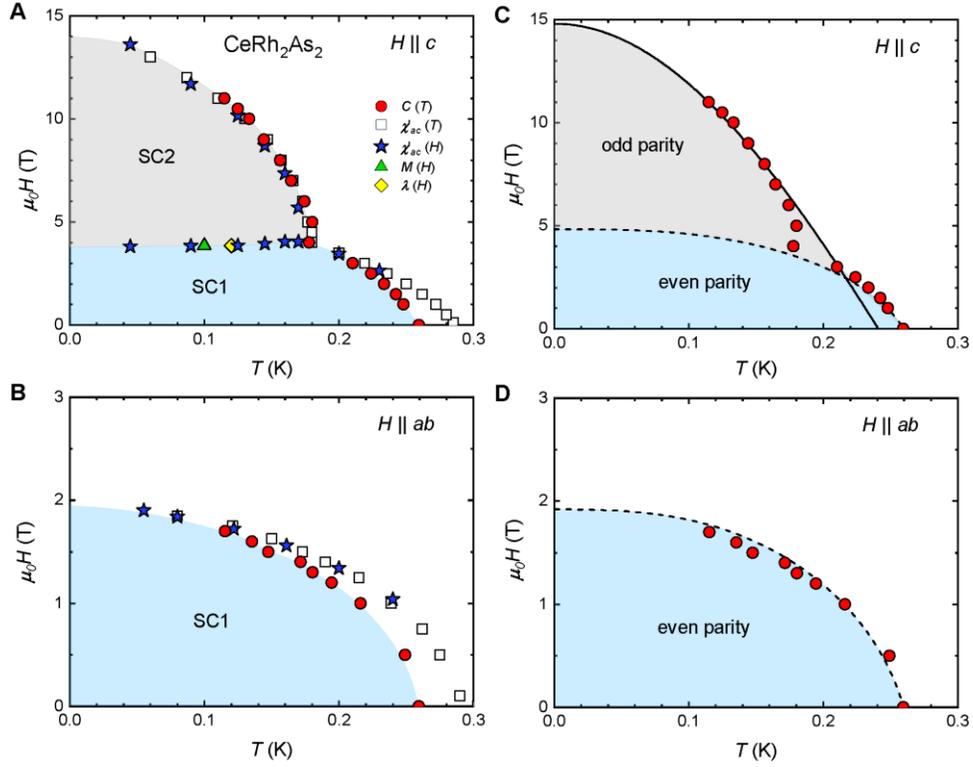

Figure 4: **Superconducting phase diagrams for CeRh$_2$As$_2$.** (**A**) $H \parallel c$ and (**B**) $H \parallel ab$. (**C**) and (**D**) Fits to the upper critical fields as explained in the text. The fit for the even parity state for $H \parallel ab$ in (**D**) is sufficient to find the critical fields for $H \parallel c$ shown in (**C**) in excellent agreement with the experimental data.

# Supplementary Materials for

# Field-induced transition from even to odd parity superconductivity in CeRh$_2$As$_2$

S. Khim,[1†*] J. F. Landaeta,[1†] J. Banda,[1] N. Bannor,[1] M. Brando,[1] P. M. R. Brydon,[2]
D. Hafner,[1] R. Küchler,[1] R. Cardoso–Gil,[1] U. Stockert,[1] A. P. Mackenzie,[1,3]
D. F. Agterberg,[4] C. Geibel,[1] and E. Hassinger[1,5*]

[1]Max Planck Institute for Chemical Physics of Solids, 01187 Dresden, Germany
[2]Department of Physics and MacDiarmid Institute for Advanced Materials and Nanotechnology,
University of Otago, P.O. Box 56, Dunedin 9054, New Zealand
[3]Scottish Universities Physics Alliance, School of Physics and Astronomy,
University of St Andrews, St Andrews KY16 9SS, United Kingdom
[4]Department of Physics, University of Wisconsin–Milwaukee,
Milwaukee, Wisconsin 53201, USA
[5]Technical University Munich, Physics department, 85748 Garching, Germany

[†]These authors contributed equally to the research.
[*]Correspondence to: elena.hassinger@cpfs.mpg.de and seunghyun.khim@cpfs.mpg.de

## Materials and Methods

Single crystal synthesis

Single crystals of $CeRh_2As_2$ were grown in Bi flux. Elemental metals with the ratio of Ce:Rh:As:Bi = 1:2:2:30 were placed in an alumina crucible which was subsequently sealed in a quartz tube filled with argon at a partial pressure of 300 mbar. The ampule was heated to 1150 ºC for 4 days and slowly cooled down to 700 ºC for a week. Grown single crystals were extracted by selectively removing the Bi flux in diluted nitric acid. The composition and the crystal structure were confirmed by Energy-dispersive x-ray spectroscopy (EDX) and x-ray diffraction measurements. Detailed crystal structure information is shown in Fig. S1, Table S1 and S2.

Specific heat

The specific heat measurements were carried out using the relaxation time method in a Quantum Design Physical Property Measurements System (PPMS) down to temperatures of 0.4 K, and a custom compensated heat-pulse calorimeter for temperatures between 0.04 and 100 K in magnetic fields up to 12 T (*43*).

Thermopower

The thermopower was measured within the *ab*-plane between 2.5 K and 290 K using a modified sample holder for the thermal transport option of a PPMS. The instrument applies a relaxation time method. A low-frequency square-wave heat pulse is generated by a resistive heater. The temperature difference along the sample is measured with two calibrated bare-chip Cernox sensors.

Ac-susceptibility

The magnetic ac-susceptibility was measured using a homemade set of compensated pick-up coils of 2 mm length and 6000 turns each. The inner and outer diameter was 1.8 mm and 5 mm, respectively. A superconducting modulation coil produced the excitation field of 175 µT at 5 Hz. The output signal of the pick-up coils was amplified using a low temperature transformer (LTT-m from CMR) with a winding ratio 1:100 and a low noise amplifier SR560 from Stanford Research Systems. Our setup uses a National Instruments 24 bits PXIe-4463 signal generator and 24 bits PXIe-4492 oscilloscope as data acquisition system with digital lock-in amplification. The ac susceptibility measurements were performed for $H \parallel ab$ and $H \parallel c$ using a small single crystal of a volume of ∼500 µm$^3$ down to 45 mK in an external magnetic field of up to 15 T in a MX400 Oxford dilution refrigerator. The data from temperature sweeps were normalized to their respective value in the normal state at 0.5 K for all the magnetic fields applied. For the field sweeps, the absolute value of the signal at different temperatures is given, normalized to the field-dependent value in the normal state at 0.35 K. Low field data were removed because of noise due to flux jumps in a superconducting magnet. The complete data is shown in Fig. S5 and S6.

Resistivity

For resistivity measurements a standard four-point method was employed with current and voltage contacts along a line perpendicular to the *c*-axis using an excitation current of 100 µA. The four contacts were spot-welded on the sample using gold wires with 25 µm diameter. The signal was amplified by a low-temperature transformer with a winding ratio of 1:100 and the output of the transformer was measured using a SR830 lock-in amplifier at a frequency of 113.7 Hz.

Magnetostriction

The linear magnetostriction coefficient $\lambda = (1/L)(dL/dH)_T$ was measured using a miniaturized high-resolution capacitance dilatometer of CuBe (*44*). The length change $\Delta L$ of the sample was measured along the *c*-axis. Using an ultrahigh-resolution Andeen-Hagerling capacitance bridge, the absolute value of the dilatometer capacitance was measured with a resolution of $10^{-6}$ pF, corresponding to a length resolution of 0.02 Å. The experiments were carried out in an Oxford Instruments K100 dilution refrigerator with a superconducting magnet in magnetic fields up to 5 T.

Magnetization

Magnetization measurements were carried out using a commercial vibration sample magnetometer, Quantum Design Magnetic Property Measurement System (MPMS-VSM), for the high-temperature range between 1.8 and 380 K, and a custom high-resolution capacitive Faraday magnetometer for the low-temperature range between 0.05 and 2 K and magnetic fields up to 12 T (*45*).

**Supplementary Text**

Subtraction of the nuclear Schottky contribution in the specific heat data at low temperatures

The low-temperature heat capacity of CeRh$_2$As$_2$ was measured in zero magnetic field and in an external field aligned along both the *c*-axis ($H \parallel c$) and the basal *ab*-plane ($H \parallel ab$) of the tetragonal crystalline structure. In all measurements we observed upturns in $C/T$ below $T \approx 0.2$ K, the intensity of which increases systematically with the square of the magnetic field. An example is shown in Fig. S2 in which the raw data for $C/T$ are plotted versus $T$ at zero field and at 10 T. At $\mu_0 H = 0$ the upturn is very weak and visible only below 0.06 K, but at $\mu_0 H \parallel ab = 10$ T (empty red circles) this effect is strong and visible below 0.2 K. With $\mu_0 H \parallel c = 10$ T (filled red circles), however, this contribution is not so distinct because it adds to that of the superconducting transition located at about 0.2 K.

These features are due to the nuclear contribution to the specific heat. Since the Ce atom has zero nuclear moment and the contribution from the Rh nuclear moment is well below 1 mK (*46, 47*), the only relevant contribution to the specific heat originates from the $^{75}$As atoms (100% abundance). $^{75}$As atoms have nuclear angular momentum $I = 3/2$, a magnetic moment $\mu_I = g_N \mu_N I$ with $\mu_N = 5.05 \times 10^{-27}$ J/T (nuclear magneton) and $g_N$, the nuclear *g*-factor, and an electric quadrupole moment $Q = 0.3 \times 10^{-28}$ m$^2$. In the tetragonal unit cell of CeRh$_2$As$_2$ they occupy two positions, As(1) and As(2) at which the field gradient $eq$ is slightly different.

In zero magnetic field the splitting of the nuclear energy levels is given by $\Delta \varepsilon = (1/2)e^2qQ$ which corresponds to a (NQR) frequency $\nu_Q = \Delta/h$. In magnetic field, however, or in magnetically ordered compounds the energy level degeneracy is further lifted and the splitting is proportional to the effective magnetic field at the nuclei $\Delta \varepsilon = -g_N \mu_N B_{\text{eff}}$ (Zeeman energy). This energy splitting $\Delta \varepsilon$ is mostly of the order of a few mK and therefore visible in specific heat data as the high-temperature part of a Schottky anomaly which is proportional to $T^{-2}$ (see, e.g., Refs. *47,48*) and is proportional to $B^2$.

The specific heat at low temperatures is then given by three contributions:

$$C = C_n + C_{el} + C_{ph} = \frac{\alpha}{T^2} + \gamma T + \beta T^3$$

, where $C_{el} + C_{ph}$ are the electronic and phonon contributions and $\alpha$ is the proportional factor in the nuclear Schottky specific heat. The phonon contribution is very small below 1 K and can safely be ignored in this analysis. Since the $C_n$ contribution is strong at low temperatures, to extract the $\alpha$ parameter it is convenient to plot the data as:

$$\frac{C(T)}{T} = \frac{\alpha}{T^3} + \gamma$$

and extrapolate the linear-in-$T$ behavior at very low $T$, i.e., in the range where the nuclear term dominates. This is exemplary shown in the inset of Fig. S2 in which $C/T$ vs $T^{-3}$ is plotted for the data at 10 T in both field directions. The data can reliably be fitted with a linear function in the low-$T$ range. The fits yield: $\alpha_{ab}(10\ T) = 3.36 \times 10^{-4}$ JK/mol and $\alpha_c(10\ T) = 3.49 \times 10^{-4}$ JK/mol. Both data sets show a similar $\alpha$ which is expected since the effective field at the As nuclei is very close to the external applied field. In fact a small magnetic anisotropy and weak Knight shifts (a few %) were observed in NQR measurements (*49*). After having extracted $\alpha$ at all fields we have plotted $\alpha$ versus $B^2$ to check consistency with the Schottky equation. A linear fit to the data yields a zero field nuclear parameter: $\alpha(0\ T) = (1.57 \pm 0.2) \times 10^{-5}$ JK/mol. We have subtracted the nuclear contribution from all data presented and analyzed in the main text. A more detailed analysis will be presented in a forthcoming paper.

Crystal electric field analysis

The crystal field Hamiltonian for Ce in the tetragonal environment is given by $H = B_2^0 O_2^0 + B_4^0 O_4^0 + B_4^4 O_4^4$ where $B_n^m$ are the crystal electric field parameters and $O_n^m$ are Stevens operators (*50, 51*). The magnetic susceptibility is calculated as

$$\chi_{CEF,i} = N_A(g_J\mu_B)^2 \frac{1}{Z} \left( \sum_{m \neq n} 2|\langle m|J_i|n\rangle|^2 \frac{1 - e^{-(\beta(E_n - E_m)}}{E_n - E_m} e^{-\beta E_n} + \sum_n |\langle n|J_i|n\rangle|^2 \beta e^{-\beta E_n} \right)$$

, where $Z = \sum_n e^{-\beta E_n}$, $\beta = 1/k_B T$, and $i = x, y, z$. The calculated inverse magnetic susceptibility including the molecular field contribution $\lambda_i$ as $\chi_i^{-1} = \chi_{CEF,i}^{-1} - \lambda_i$ is given to fit the experimental data. The presented fits (see Fig. S3) are obtained from the doublet ground state $|\Gamma_7^{(1)}\rangle = 0.88 |\mp 3/2\rangle - 0.47 |\pm 5/2\rangle$ with the first excited state $|\Gamma_6\rangle = |\pm 1/2\rangle$ at 30 K and the second excited level $|\Gamma_7^{(2)}\rangle = 0.47 |\pm 3/2\rangle + 0.88 |\mp 5/2\rangle$ at 180 K. $\lambda_i$ is set to 36 mol/emu for both field. These states are obtained from a diagonalization of the crystal field Hamiltonian with the parameters of $B_2^0 = 6.5$ K, $B_4^0 = 0.1$ K, and $B_4^4 = 2.8$ K. The energy splitting for the two lower-lying states is comparable to the Kondo energy scale of $T_K \sim 20 - 40$ K, which is given by the relation of $S_{mag}(T_K/2) = (1/2)R \ln x$, where $x = 2$ for a doublet and 4 for a quartet ground state. The ground state can have a nature of a quartet state when the wave function of the first-excited level partially contributes to the ground state through a possible mixing channel such as Kondo interactions.

Determination of the g-factor

The anisotropic g-factors used to estimate the Pauli limiting fields were obtained from the dc magnetic susceptibility ($\chi_S$) and the specific heat as follow. For the spin-1/2 ($J = 1/2$) single-ion Kondo system, the Wilson ratio written as

$$R_W = \frac{\pi^2 k_B^2}{\mu_0 (\mu_{eff})^2} \frac{\chi_s}{\gamma}$$

, where $(\mu_{eff})^2 = g^2 \mu_B^2 J(J+1)$ and $\mu_0$ is the magnetic permeability of vacuum, is given to be 2 (Ref. 52). The Wilson ratio for the Kondo-lattice CeRh$_2$As$_2$ system is expected to be close to this value as the system does not include strong spin-spin interactions. The magnetic susceptibility ($\chi_s$) for $H \parallel c$ is almost saturated below 10 K and reached to ~ $8.4 \times 10^{-3}$ emu/mol. For $H \parallel ab$, the $\chi_s$ tends to saturate upon cooling but diverges in the zero temperature limit. The saturation value is estimated to be $1.4 \times 10^{-2}$ emu/mol. The Sommerfeld coefficient ($\gamma$) is estimated to be ~ 1 J/molK$^2$ from the specific heat data at high fields in the zero temperature limit where superconductivity completely vanishes for $H \parallel ab$. From the assumed value of $R_W = 2$, the effective magnetic moments are deduced to be 1.24 $\mu_B$ (0.96 $\mu_B$) for $H \parallel ab$ ($H \parallel c$). Accordingly, the saturated moments ($\mu_{sat} = \mu_{eff}/\sqrt{3}$) are 0.71 $\mu_B$ (0.55 $\mu_B$) and the resultant g-factors are 1.43 and 1.11 for $H \parallel ab$ and $H \parallel c$, respectively.

The $\mu_{sat}$ can be alternatively estimated from the wavefunction of the ground state of the localized f-orbital crystal electric field configuration of Ce, as described in the above section. These values are 0.790 $\mu_B$ and 0.543 $\mu_B$ for $H \parallel ab$ and $H \parallel c$, respectively, consistent with the values from the $R_W$.

Estimation of the Maki parameter from normal state properties

The Maki parameter obtained from the fit agrees with an estimation in the dirty limit from normal state properties $\alpha_m = 2e^2\hbar\gamma\rho_n / 2m_e\pi^2 k_B^2 V_{mol} = 5.7$ (Ref. 21), with the electronic charge $e$, Planck's constant divided by $2\pi$ $\hbar$, the specific-heat Sommerfeld coefficient just above the superconducting state $\gamma = 1$ J/molK$^2$, the normal state resistivity of samples of the same batch $\rho_n = 13.3$ $\mu\Omega$cm, the electron mass $m_e$, the Boltzmann constant $k_B$, and the molar volume $V_{mol} = 545 \times 10^{-7}$ m$^3$/mol. A large Maki parameter usually characterizes systems which are susceptible to going into a state with a modulated SC order parameter, i.e. a Fulde-Ferrell-Larkin-Ovchinnikov (FFLO) state, if they are clean enough (53, 54). For our samples with a relatively large $\rho_0$ this is unlikely and the kink feature is not typical for an FFLO state where a smooth upturn is expected. Note that we don't detect the helical phase expected for in-plane fields (24).

Relation between slopes of transition lines and specific heat jump at a tricritical point.

In the following, we assume that the lines separating the superconducting states 1 and 2 from the normal state N (lines 1-N and 2-N, respectively) are second order and the line separating the two superconducting states (line 1-2) is first order as shown in the schematic phase diagram (FIG. S7). Following the literature (34), the slopes of the transition lines near the multicritical point $p_1 = (dH/dT)_{1-N}$, $p_2 = (dH/dT)_{2-N}$, and $p_3 = (dH/dT)_{1-2}$ are related by

$$p_3 = p_1 \left(\frac{r-1}{r-y}\right)$$

where $r$ is given by the squareroot of the ratio of the specific heat jumps $r = (\Delta C_{1-N}/\Delta C_{2-N})^{1/2}$, and $y = |p_1|/|p_2| < 1$. We extract the slopes from the susceptibility measurements, since the relative error bar between two measurements at constant field or constant temperature is lowest there. We find experimentally $p_1 = (-25 \pm 5)$ T/K, $p_2 < -1000$ T/K and $p_3 = (2 \pm 1)$ T/K. Since $|p_1| \ll |p_2|$ this implies that $y \approx 0$ and $p_3/p_1 \approx -0.08 = (r - 1)/r$ so that $r \approx 0.92$. Hence the specific heat jump $\Delta C_{1\text{-}N} = (0.85 \pm 0.1)\, \Delta C_{2\text{-}N}$ where $\Delta C_{1\text{-}N} = \Delta C(H < H_{\text{kink}})$ and $\Delta C_{2\text{-}N} = \Delta C(H > H_{\text{kink}})$ (See Fig. S4).

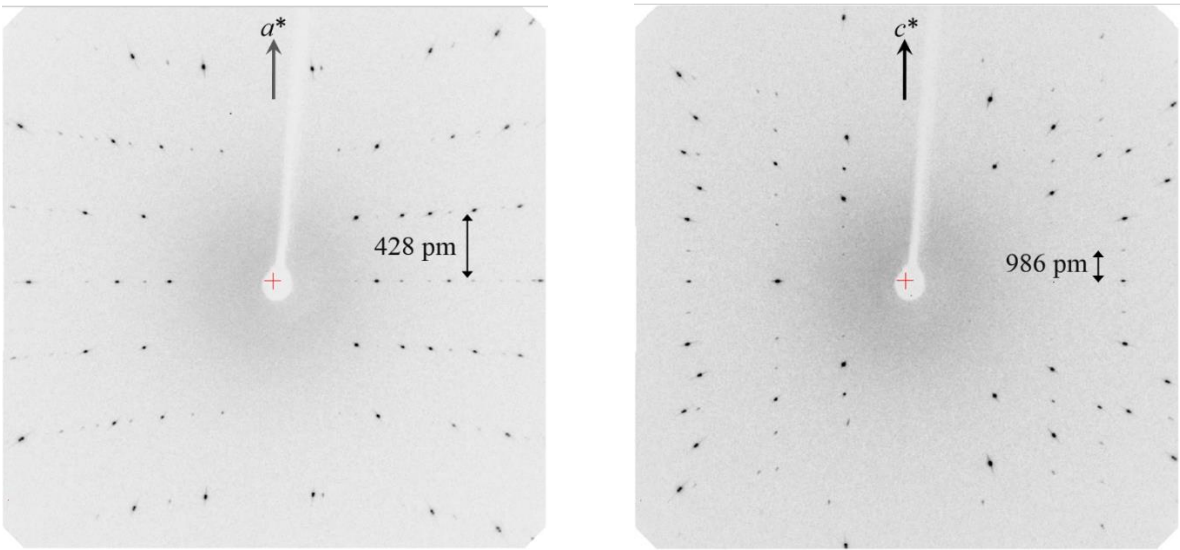

**Fig. S1.**
CeRh$_2$As$_2$ single crystal axial images (Mo Kα radiation).

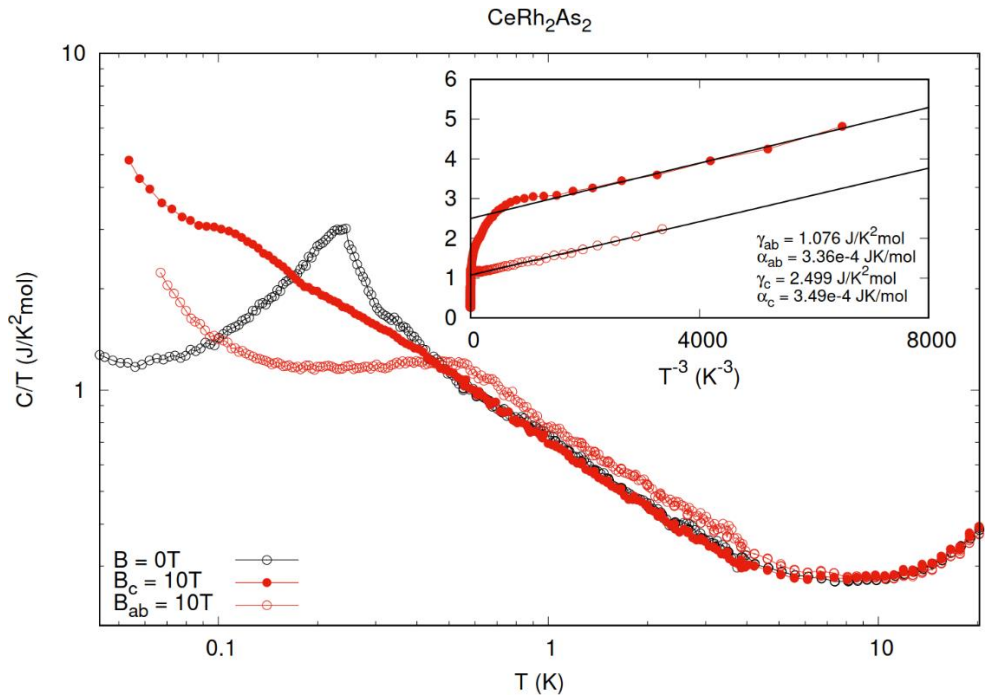

**Fig. S2.**
Specific heat of CeRh$_2$As$_2$ measured in zero magnetic field and in a field of 10 T aligned along both the crystallographic *c*-axis (filled red circles) and the *ab*-plane (empty red circles). Inset: Plot of $C/T$ vs $T^{-3}$ with linear fits for the lowest temperature data.

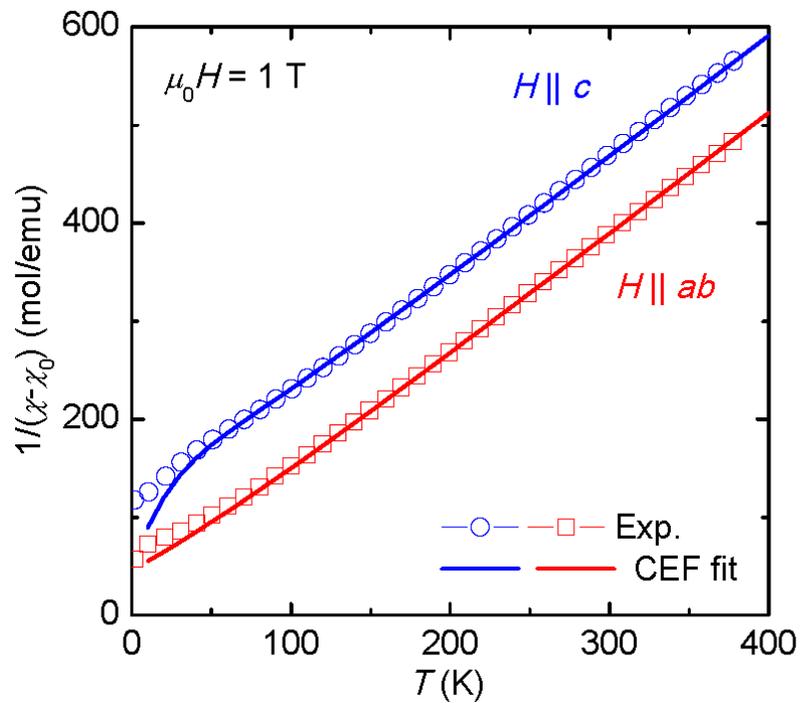

**Fig. S3.**
Inverse magnetic susceptibility in single crystalline $CeRh_2As_2$ and the model fits for the crystal electric field configuration of Ce (see the supplementary text).

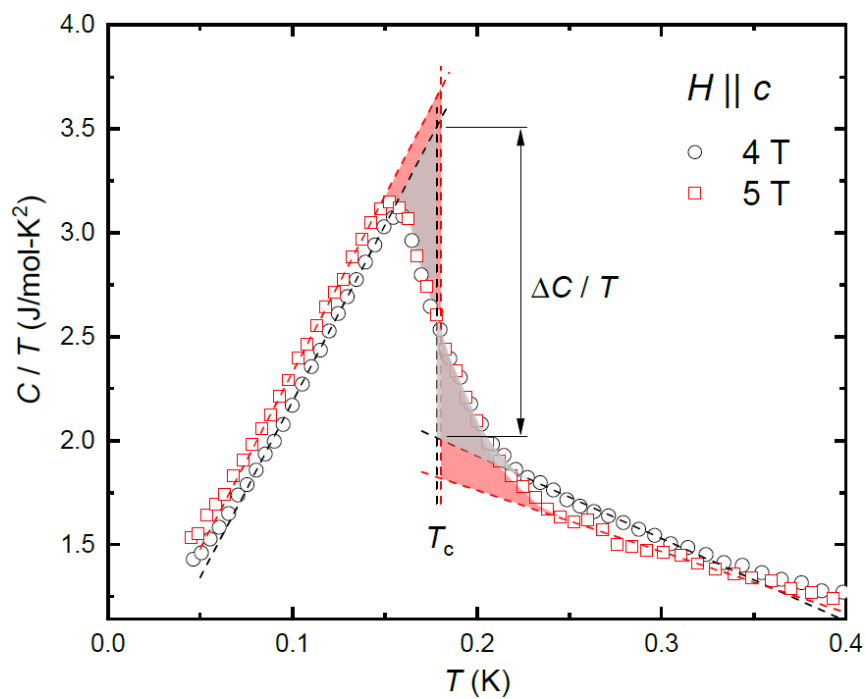

**Fig. S4.**
$C/T$ data in 4 and 5 T for $H \parallel c$ and the example of the determination of a thermodynamic bulk $T_c$ from the entropy balance. $T_c$ is defined at the midpoint of the $C/T$ jump to lead to the same areas enclosed by the experimental data and the linearly interpolated lines below and above the jump to $T_c$. The $\Delta C/T$ value is enhanced from 1.49 J/molK$^2$ to 1.87 J/molK$^2$ when the field increases from 4 to 5 T. From these data, the change of the jump can be estimated to $\Delta C(4\text{T})/\Delta C(5\text{T}) \approx 0.8$, in agreement with the expected change calculated from the slopes of the transition lines in the phase diagram (see the supplementary text).

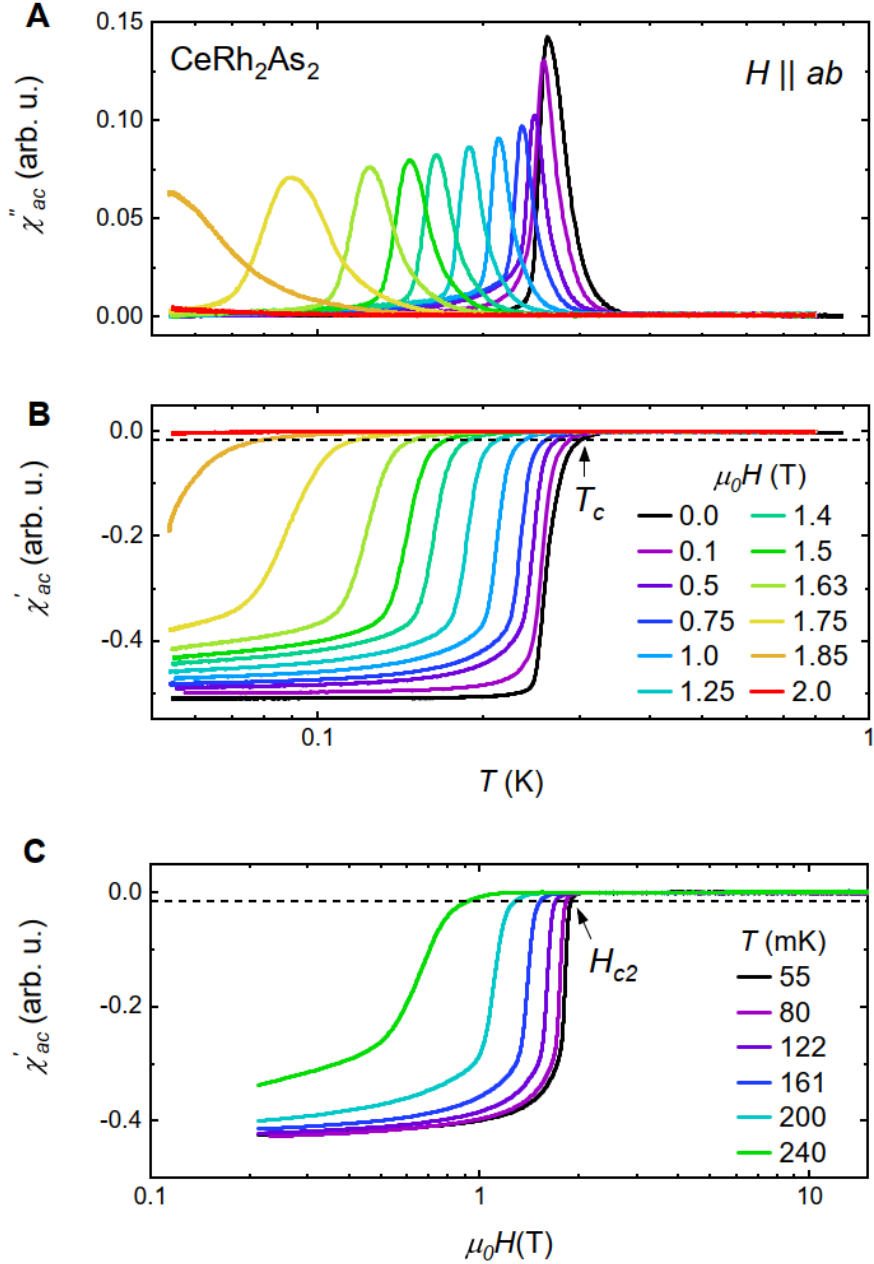

**Fig. S5.**
Complete data set of magnetic ac-susceptibility in $H \parallel ab$. **A,B**, The temperature dependence of the imaginary and real part signal in various magnetic fields. **C**, Field dependence of the real part signal in various temperatures. The horizontal dotted lines in (**B**) and (**C**) denote the criterion to choose $T_c$ and $H_{c2}$, respectively.

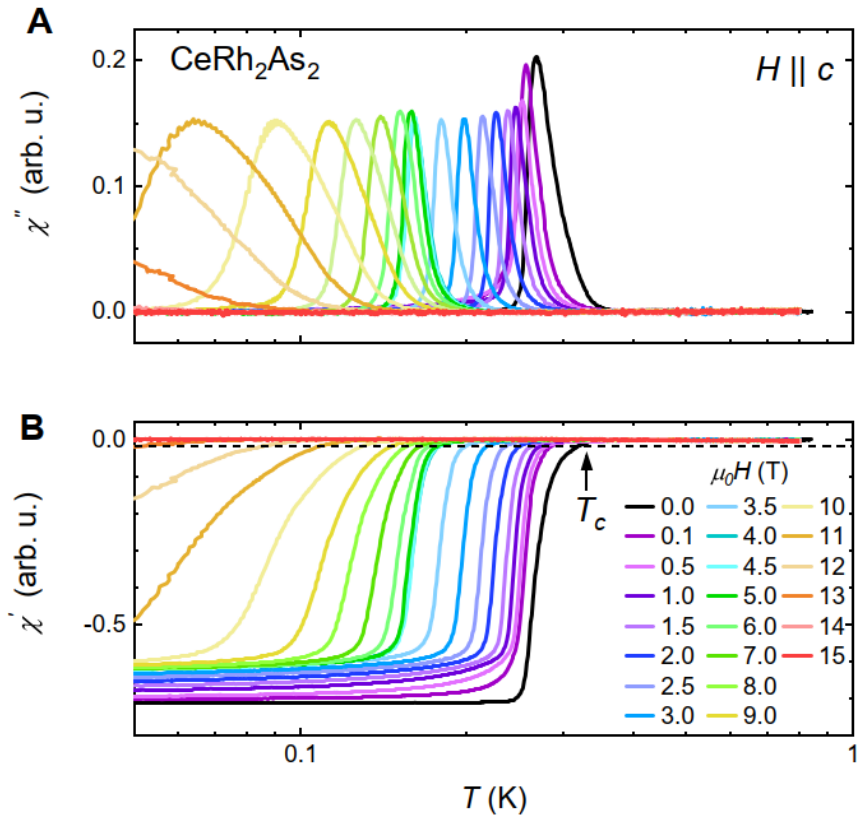

**Fig. S6.**
Complete data set of magnetic ac susceptibility in $H \parallel c$. **A,B,** The temperature dependence of the imaginary and real part signal in various magnetic fields. The horizontal dotted line in (**B**) denote the criterion to choose $T_c$.

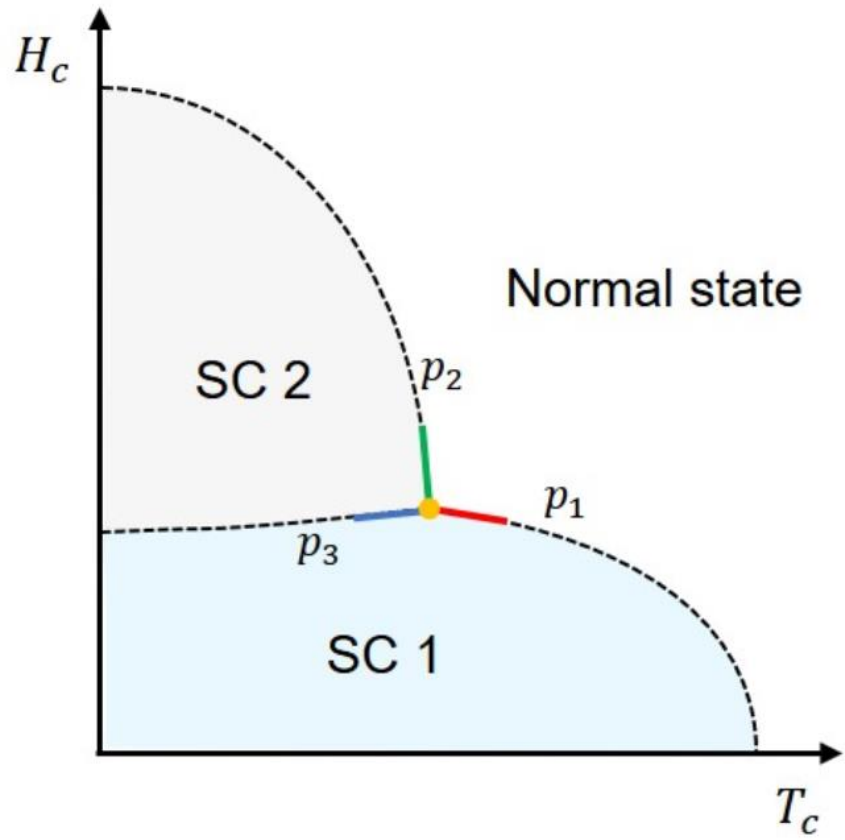

**FIG. S7.**

Schematic phase diagram for $H \parallel c$. Here we show the relationship of the slopes between SC1, SC2, and the normal state.

**Table S1.**

Crystallographic data and details of the crystal structure analysis

| | |
|---|---|
| Formula; molar mass | $CeRh_2As_2$; 495.78 amu |
| Crystal | Metallic prism; $0.076 \times 0.071 \times 0.047$ mm$^3$ |
| Space group; formula units | $P4/nmm$ (No.129, origin at center); $Z = 2$ |
| Lattice parameter (294 K)* | $a = 428.01(2)$ pm |
| | $c = 986.16(5)$ pm |
| Volume; density | $180.66(2) \times 10^6$ pm$^3$; 9.114 gcm$^{-3}$ |
| Data collection | Rigaku AFC7 diffraction system, Saturn 724+ detector |
| | Mo K$\alpha$ radiation $\lambda = 71.073$ pm, graphite monochromator |
| | 900 exposures, $\Delta\varphi = 0.8°$ |
| Measured Range | $2.1° \leq \theta \leq 43.0°$ |
| Structure refinement | SHELXL-97 (*55*) as implemented in WinGX (*56*) |
| | Full-matrix least-squares on $F^2$ (15 parameter) |
| Absorption correction | Numerical ($\mu = 39.26$ mm$^{-1}$) |
| Max. and min. transmission | 1.00 and 0.32 |
| Measured/unique reflections | 3756 / 438 |
| $R_{int}$ | 0.048 |
| Observed reflections (Fo > 4.0 σ(Fo)) | 401 |
| $R(F)$; $wR2$ | 0.030; 0.067 |
| Goof = S = | 1.077 |
| $\Delta\rho_{max}$ | 3.83 e Å$^{-3}$ |
| $\Delta\rho_{min}$ | – 5.92 e Å$^{-3}$ |

*Refined on 40 reflections from powder diffraction data using the WinCSD software package (*57*) and LaB$_6$ as internal standard ($a = 415.69(1)$ pm)

**Table S2.**

Atomic coordinates and anisotropic displacement parameters (ADP) $U_{ij}$ (pm$^2$) for CeRh$_2$As$_2$. $U_{eq}$ is defined as one third of the orthogonalized $U_{ij}$ tensor. The anisotropic displacement factor exponent takes the form: $[-2\pi^2(h^2 a^{*2} U_{11} +...+ 2hk a^* b^* U_{12})]$. $U_{23} = U_{13} = U_{12} = 0$

| Atom  | Site | x   | y   | z          | $U_{eq}$ | $U_{11}$ | $U_{22}$ | $U_{33}$ |
|-------|------|-----|-----|------------|----------|----------|----------|----------|
| Ce(1) | 2c   | ¼   | ¼   | 0.25469(3) | 67(1)    | 72(1)    | $U_{11}$ | 57(2)    |
| Rh(1) | 2a   | ¾   | ¼   | 0          | 76(1)    | 87(2)    | $U_{11}$ | 55(2)    |
| Rh(2) | 2c   | ¼   | ¼   | 0.61742(4) | 72(1)    | 78(1)    | $U_{11}$ | 60(2)    |
| As(1) | 2a   | ¾   | ¼   | 0.5        | 70(2)    | 70(2)    | $U_{11}$ | 72(3)    |
| As(2) | 2c   | ¼   | ¼   | 0.86407(7) | 66(1)    | 73(2)    | $U_{11}$ | 51(2)    |